\documentclass{article}

\usepackage{arxiv}

\usepackage[utf8]{inputenc}
\usepackage[T1]{fontenc}
\usepackage{hyperref}
\usepackage{url}
\usepackage{booktabs}
\usepackage{amsfonts}
\usepackage{amssymb}
\usepackage{amsmath}
\usepackage{microtype}
\usepackage{graphicx}
\usepackage[numbers]{natbib}
\usepackage{doi}
\usepackage{xcolor}
\usepackage{listings}
\usepackage{fancyvrb}
\usepackage{tikz}
\usetikzlibrary{shapes.geometric, arrows.meta, positioning, fit, calc, backgrounds, decorations.pathreplacing}
\usepackage{enumitem}
\usepackage{float}

\definecolor{codebg}{gray}{0.96}
\definecolor{codegreen}{rgb}{0.0,0.5,0.0}
\definecolor{codegray}{rgb}{0.5,0.5,0.5}
\definecolor{codepurple}{rgb}{0.58,0,0.82}

\lstdefinestyle{pythonstyle}{
  backgroundcolor=\color{codebg},
  commentstyle=\color{codegreen},
  keywordstyle=\color{blue},
  numberstyle=\tiny\color{codegray},
  stringstyle=\color{codepurple},
  basicstyle=\ttfamily\footnotesize,
  breaklines=true,
  frame=single,
  language=Python,
  numbers=left,
  numbersep=5pt,
  showstringspaces=false,
  tabsize=2
}

\title{SQL Query Engine: A Self-Healing LLM Pipeline for Natural Language to PostgreSQL Translation}

\author{
  Muhammad Adeel Ijaz \\
  \texttt{code.adeel@gmail.com} \\
  \url{https://github.com/codeadeel/sqlqueryengine}
}

\date{}

\renewcommand{\headeright}{Technical Report}
\renewcommand{\undertitle}{Technical Report}
\renewcommand{\shorttitle}{SQL Query Engine: Self-Healing LLM Pipeline}

\hypersetup{
  pdftitle={SQL Query Engine: A Self-Healing LLM Pipeline for Natural Language to PostgreSQL Translation},
  pdfsubject={cs.DB, cs.CL, cs.AI},
  pdfauthor={Muhammad Adeel Ijaz},
  pdfkeywords={text-to-SQL, large language models, self-healing, PostgreSQL, natural language interfaces},
}

\begin{document}
\maketitle

\begin{abstract}
We present \textbf{SQL Query Engine}, an open-source, self-hosted service that translates natural language questions into validated PostgreSQL queries through a two-stage pipeline driven by large language models (LLMs). The first stage performs automatic schema introspection and LLM-guided SQL generation; a multi-strategy response parser extracts SQL from any LLM output format (JSON, code blocks, or raw text) without requiring structured output APIs or function calling. The second stage executes the generated query against a live PostgreSQL instance and, upon failure or empty results, enters an iterative \emph{self-healing} loop in which the LLM diagnoses the error, using full SQLSTATE codes and PostgreSQL diagnostic messages, and produces a corrected query. Two mechanisms prevent regressions: \emph{early-accept} returns successful queries immediately without LLM re-evaluation, and \emph{best-result tracking} preserves the best partial result across retries. The system caches schema context per session in Redis, streams real-time progress events via Redis Pub/Sub and Server-Sent Events (SSE), and exposes both a native REST API and a fully OpenAI-compatible \texttt{/v1/chat/completions} endpoint, so existing tools such as Open~WebUI and the OpenAI Python SDK work without modification. All database connections are enforced as read-only at the driver level. We evaluate the pipeline on two fronts: (1)~a \emph{synthetic} ablation study across five LLM backends and 75 gold-standard questions spanning three purpose-built PostgreSQL databases, where the self-healing loop yields up to +9.3 percentage-point accuracy gains with zero regressions on the best model (Llama~4~Scout~17B, 57.3\%); and (2)~the \emph{BIRD} benchmark, a large-scale real-world evaluation on 437 questions across 11 databases migrated from SQLite to PostgreSQL, where the full pipeline reaches 49.0\% execution accuracy (GPT-OSS-120B) with a +4.6 percentage-point self-healing improvement. Source code is publicly available at \url{https://github.com/codeadeel/sqlqueryengine}.
\end{abstract}


\section{Introduction}
\label{sec:introduction}

Translating natural language into executable SQL queries, commonly referred to as the \emph{text-to-SQL} problem, has long been a research objective in the database and natural language processing communities~\citep{zhong2017seq2sql, yu2018spider}. The emergence of large language models (LLMs) such as GPT-4~\citep{openai2023gpt4} and open-weight alternatives like LLaMA~\citep{touvron2023llama} has substantially advanced the state of the art, with recent methods achieving execution accuracies above 85\% on the Spider benchmark~\citep{pourreza2023dinsql, gao2024dailsql}. Yet a considerable gap remains between benchmark performance and reliable deployment in production settings, where schemas are large, dynamic, and undocumented.

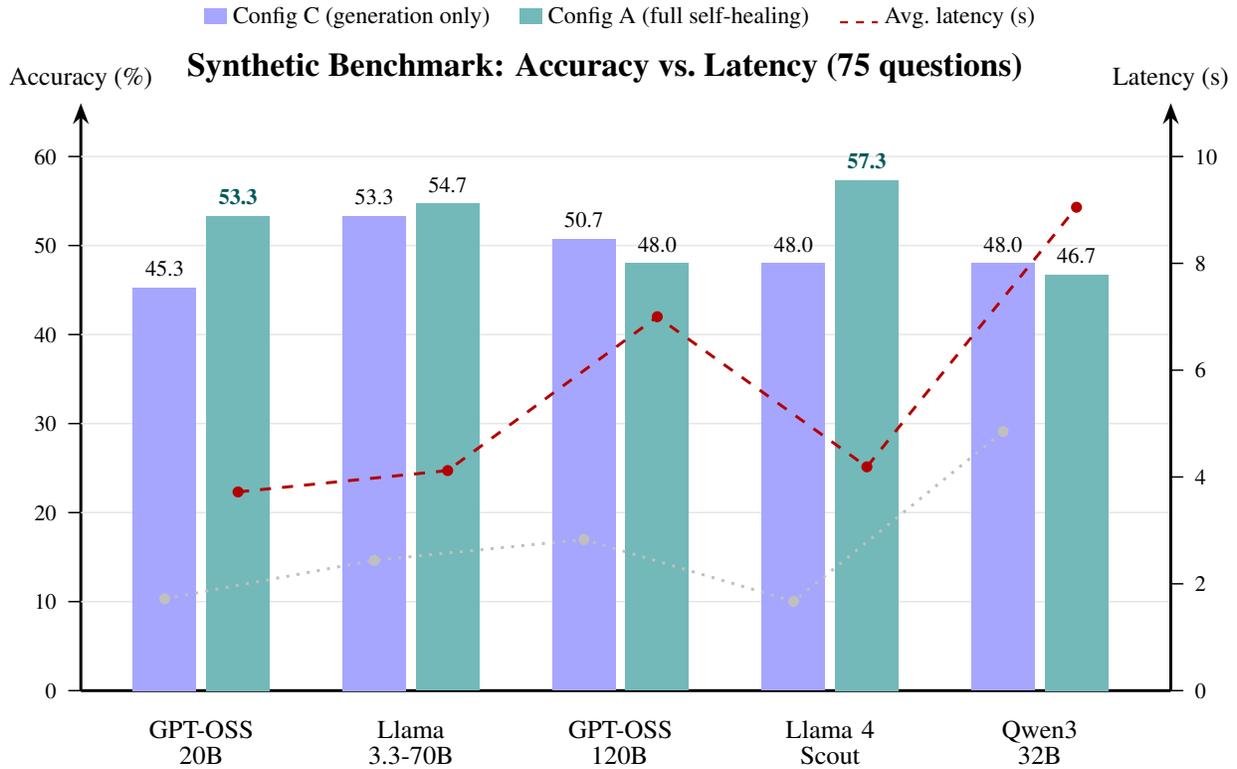
\begin{figure}[ht]
\centering

\small
\tikz\fill[blue!35] (0,0) rectangle (0.3,0.2); Config~C (generation only)
\quad
\tikz\fill[teal!55] (0,0) rectangle (0.3,0.2); Config~A (full self-healing)
\quad
\tikz\draw[red!70!black, thick, dashed] (0,0.1) -- (0.5,0.1); Avg.\ latency (s)

\vspace{4pt}
\resizebox{\textwidth}{!}{%
\begin{tikzpicture}[
  x=2.0cm, y=0.085cm,
]

  \node[font=\small\bfseries] at (2.5,70) {Synthetic Benchmark: Accuracy vs.\ Latency (75 questions)};

  \draw[thick, -{Stealth[length=2mm]}] (0,0) -- (0,66) node[above, font=\scriptsize] {Accuracy (\%)};
  \foreach \y in {0,10,20,30,40,50,60} {
    \draw[gray!20] (0,\y) -- (5.2,\y);
    \draw (0,\y) -- (-0.06,\y) node[left, font=\tiny] {\y};
  }

  \draw[thick, -{Stealth[length=2mm]}] (5.2,0) -- (5.2,66) node[above, font=\scriptsize] {Latency (s)};
  \foreach \l/\v in {0/0, 2/12, 4/24, 6/36, 8/48, 10/60} {
    \draw (5.2,\v) -- (5.26,\v) node[right, font=\tiny] {\l};
  }

  \draw[thick] (0,0) -- (5.2,0);

  \fill[blue!35] (0.25,0) rectangle (0.55,45.3);
  \fill[teal!55] (0.60,0) rectangle (0.90,53.3);
  \node[font=\tiny, above] at (0.40,45.3) {45.3};
  \node[font=\tiny, above, text=teal!70!black, font=\tiny\bfseries] at (0.75,53.3) {53.3};
  \node[font=\scriptsize, below, text width=1.4cm, align=center] at (0.575,-2) {GPT-OSS\\[-1pt]20B};

  \fill[blue!35] (1.25,0) rectangle (1.55,53.3);
  \fill[teal!55] (1.60,0) rectangle (1.90,54.7);
  \node[font=\tiny, above] at (1.40,53.3) {53.3};
  \node[font=\tiny, above] at (1.75,54.7) {54.7};
  \node[font=\scriptsize, below, text width=1.4cm, align=center] at (1.575,-2) {Llama\\[-1pt]3.3-70B};

  \fill[blue!35] (2.25,0) rectangle (2.55,50.7);
  \fill[teal!55] (2.60,0) rectangle (2.90,48.0);
  \node[font=\tiny, above] at (2.40,50.7) {50.7};
  \node[font=\tiny, above] at (2.75,48.0) {48.0};
  \node[font=\scriptsize, below, text width=1.4cm, align=center] at (2.575,-2) {GPT-OSS\\[-1pt]120B};

  \fill[blue!35] (3.25,0) rectangle (3.55,48.0);
  \fill[teal!55] (3.60,0) rectangle (3.90,57.3);
  \node[font=\tiny, above] at (3.40,48.0) {48.0};
  \node[font=\tiny, above, text=teal!70!black, font=\tiny\bfseries] at (3.75,57.3) {57.3};
  \node[font=\scriptsize, below, text width=1.4cm, align=center] at (3.575,-2) {Llama~4\\[-1pt]Scout};

  \fill[blue!35] (4.25,0) rectangle (4.55,48.0);
  \fill[teal!55] (4.60,0) rectangle (4.90,46.7);
  \node[font=\tiny, above] at (4.40,48.0) {48.0};
  \node[font=\tiny, above] at (4.75,46.7) {46.7};
  \node[font=\scriptsize, below, text width=1.4cm, align=center] at (4.575,-2) {Qwen3\\[-1pt]32B};

  \draw[gray!50, thick, dotted]
    (0.40,10.32) -- (1.40,14.64) -- (2.40,16.98) -- (3.40,10.02) -- (4.40,29.10);
  \foreach \x/\y in {0.40/10.32, 1.40/14.64, 2.40/16.98, 3.40/10.02, 4.40/29.10} {
    \fill[gray!50] (\x,\y) circle (1.5pt);
  }

  \draw[red!70!black, thick, dashed]
    (0.75,22.32) -- (1.75,24.72) -- (2.75,42.00) -- (3.75,25.14) -- (4.75,54.30);
  \foreach \x/\y in {0.75/22.32, 1.75/24.72, 2.75/42.00, 3.75/25.14, 4.75/54.30} {
    \fill[red!70!black] (\x,\y) circle (1.5pt);
  }

\end{tikzpicture}%
}
\caption{Synthetic benchmark: execution accuracy (bars, left axis) and average query latency (lines, right axis) before and after self-healing. Scout achieves the highest accuracy (57.3\%, +9.3pp) with moderate latency increase (1.67s $\to$ 4.19s). The latency cost of self-healing scales with the number of repair iterations needed.}
\label{fig:intro_comparison}
\end{figure}

A core limitation of single-pass text-to-SQL approaches is their inability to recover from errors. When an LLM generates an incorrect query, most systems either return the raw database error to the user or silently give up~\citep{shi2024survey}. In practice, however, many of these failures are \emph{recoverable}: the query may contain a misspelled column name, an incorrect type cast, or an overly restrictive filter that a second LLM call could fix if provided with the right diagnostic information. This observation motivates the design of \textbf{SQL Query Engine}, an open-source, self-hosted pipeline that pairs LLM-driven SQL generation with an execution-grounded \emph{self-healing} loop. Figure~\ref{fig:intro_comparison} summarizes the key result: across two independent evaluation suites and five LLM backends, the self-healing loop consistently recovers failed queries and raises execution accuracy by up to +9.3 percentage points on synthetic benchmarks and +4.6 percentage points on BIRD~\citep{li2024bird}, a large-scale real-world benchmark.

Beyond the raw accuracy numbers, three production-oriented challenges motivate the design. First, real-world database schemas are often large and evolve over time; the system must introspect the schema dynamically rather than relying on a static, curated representation. Second, LLM-generated queries frequently fail on the first attempt due to schema mismatches, incorrect type casts, or overly restrictive filters, yet most existing systems either surface the error to the user or silently give up. Third, enterprise environments demand safety guarantees (e.g., that the system cannot modify data) and observability into the query repair process.

SQL Query Engine addresses these concerns through three principal design decisions:

\begin{enumerate}[leftmargin=1.5em]
  \item A \textbf{two-stage pipeline} that separates SQL generation from SQL evaluation and repair, with each stage implemented as a composable module exposable via independent API endpoints.
  \item A \textbf{self-healing loop} in the evaluation stage that captures full PostgreSQL error diagnostics (SQLSTATE codes, diagnostic messages, hints, and Python tracebacks) and feeds them back to the LLM for iterative correction, guarded by \emph{early-accept} and \emph{best-result tracking} to prevent regressions.
  \item A \textbf{session-aware caching layer} backed by Redis that stores per-user schema descriptions and conversation history, eliminating redundant database introspection and enabling multi-turn interactions.
\end{enumerate}

The system is LLM-agnostic: it communicates with any OpenAI-compatible endpoint (Ollama, vLLM, OpenAI, LiteLLM, and others), so switching between local open-weight models and hosted APIs requires no code changes. It also exposes an OpenAI-compatible API surface itself, so clients such as Open~WebUI~\citep{openwebui2024} or the OpenAI Python SDK can connect to it transparently.

The remainder of this report is organized as follows. Section~\ref{sec:related} surveys related work. Section~\ref{sec:architecture} details the system architecture. Section~\ref{sec:selfhealing} describes the self-healing algorithm. Section~\ref{sec:streaming} covers the real-time streaming protocol. Section~\ref{sec:api} describes the dual API surface. Section~\ref{sec:safety} discusses safety mechanisms. Section~\ref{sec:evaluation} presents empirical evaluations on both synthetic and BIRD benchmarks across five LLM backends. Section~\ref{sec:discussion} reflects on design trade-offs. Section~\ref{sec:conclusion} concludes.

\section{Related Work}
\label{sec:related}

\paragraph{Text-to-SQL with LLMs.}
The text-to-SQL field has evolved from sequence-to-sequence models~\citep{zhong2017seq2sql} and relation-aware encoders~\citep{wang2020ratsql} toward LLM-based pipelines that use in-context learning and chain-of-thought prompting~\citep{brown2020language, wei2022chain}. DIN-SQL~\citep{pourreza2023dinsql} decomposes the task into schema linking, classification, SQL generation, and self-correction sub-tasks, achieving 85.3\% execution accuracy on Spider; by breaking the problem into smaller, more tractable sub-problems, DIN-SQL shows that task decomposition can substantially reduce the cognitive load on the LLM at each step. DAIL-SQL~\citep{gao2024dailsql} investigates prompt engineering strategies (question representation, example selection, and example organization) and reaches 86.6\% through a combination of supervised fine-tuning and in-context learning, establishing that the choice of prompt format matters as much as the choice of model. C3~\citep{dong2023c3} demonstrates that a carefully designed zero-shot prompt with ChatGPT can reach 82.3\% without any training examples, suggesting that strong base models may not need in-context demonstrations for moderately complex queries. CHASE-SQL~\citep{chasesql2024} takes a different approach by employing multi-path reasoning: it generates candidate SQL queries through several strategies (divide-and-conquer decomposition, chain-of-thought with execution plans, and instance-aware synthetic examples) and then selects the best candidate via a preference-optimized ranking agent, achieving 73.0\% on the BIRD test set. Recent surveys~\citep{shi2024survey, hong2024nextgen} provide comprehensive taxonomies of these approaches, identifying prompt engineering and fine-tuning as the two dominant paradigms.

\paragraph{Multi-agent and decomposed approaches.}
A growing line of work distributes the text-to-SQL task across multiple specialized agents. MAC-SQL~\citep{macsql2023} introduces a three-agent framework comprising a \emph{Selector} that prunes large schemas to the relevant subset, a \emph{Decomposer} that breaks complex questions into sub-queries via few-shot chain-of-thought, and a \emph{Refiner} that iteratively validates and corrects the generated SQL, reaching 59.6\% on BIRD and 86.8\% on Spider. DTS-SQL~\citep{pourreza2024dtssql} decomposes the pipeline into two fine-tuning stages (schema linking followed by SQL generation) and shows that a 7B-parameter DeepSeek model can match proprietary LLM performance (84.4\% on Spider) when each stage is trained independently on its specific sub-task. CodeS~\citep{li2024codes} builds a family of open-source models (1B--15B parameters) by incrementally pre-training StarCoder on SQL-centric corpora, demonstrating that targeted domain pre-training enables small models to compete with GPT-4 on text-to-SQL while preserving data privacy through local deployment. Our system achieves a similar separation of concerns through its two-stage pipeline but uses a single LLM instance with different prompts and Pydantic validation schemas for each stage, reducing infrastructure complexity while remaining compatible with any of these models as a backend.

\paragraph{Self-correction and iterative repair.}
The idea of using execution feedback to improve LLM-generated code has gained traction across multiple domains. \citet{chen2024teaching} show that teaching LLMs to ``self-debug'' by feeding back execution traces and error messages substantially improves code generation accuracy; their key insight is that the error message itself often contains enough information for the model to identify and fix its mistake without external guidance. In the text-to-SQL domain specifically, ReFoRCE~\citep{deng2025reforce} performs iterative self-refinement with bounded retries, combining pattern-based table grouping for schema compression with LLM-guided schema linking and dialect-aware syntax correction; it currently leads the Spider~2.0 leaderboard. MAGIC~\citep{askari2025magic} takes a complementary approach by using an LLM to \emph{generate} self-correction guidelines from training data rather than hand-authoring them, achieving correction quality that surpasses manually written heuristics on both Spider and BIRD. Our self-healing loop follows a similar philosophy but integrates tightly with PostgreSQL diagnostics, feeding the full SQLSTATE error taxonomy and diagnostic hints back to the LLM in a structured evaluation prompt. Unlike CHASE-SQL's multi-path generation strategy, our system repairs a single query path iteratively, which keeps latency bounded and avoids the cost of generating multiple candidate queries in parallel.

\paragraph{Benchmarks.}
Spider~\citep{yu2018spider} has been the dominant cross-domain text-to-SQL benchmark since 2018, comprising 10,181 questions across 200 databases with complex SQL constructs including joins, nested queries, and set operations. BIRD~\citep{li2024bird} was introduced to address Spider's limitations by focusing on large-scale, real-world databases: it contains 12,751 question--SQL pairs across 95 databases totaling 33.4\,GB of data in 37 professional domains. A key difference is that BIRD emphasizes database \emph{content} understanding---the ability to reason about actual data values, messy formats, and domain-specific conventions---rather than schema structure alone. Even the best models achieve only around 40\% execution accuracy on BIRD's development set when evaluated without oracle knowledge evidence, compared to over 85\% on Spider, highlighting a substantial gap between synthetic benchmark performance and real-world applicability. Our evaluation uses the BIRD mini-dev split (500 questions, 11 databases) migrated from SQLite to PostgreSQL to match our system's target dialect.

\paragraph{Robust response parsing.}
Ensuring reliable extraction of SQL from LLM output is a practical challenge. PICARD~\citep{scholak2021picard} constrains auto-regressive decoding at the token level to guarantee valid SQL syntax, but requires access to the model's logits during generation. Structured output APIs (e.g., LangChain's \texttt{with\_structured\_output}) enforce JSON schemas but require model support and fail with models that produce free-form text. Our approach instead uses a \emph{multi-strategy response parser}: a five-strategy cascade that attempts JSON parsing, embedded JSON extraction, code block extraction, \texttt{SELECT} regex matching, and raw text fallback. This design works with any model regardless of its output format, including reasoning models (Qwen3, DeepSeek-R1) whose \texttt{<think>} tags are stripped before parsing.

\paragraph{Open-source text-to-SQL systems.}
Several open-source projects occupy this space. Vanna provides a retrieval-augmented generation approach to text-to-SQL with support for multiple databases and LLM backends, training a RAG model on DDL statements and documentation to improve generation quality. DB-GPT achieves 82.5\% on Spider after fine-tuning and supports multiple fine-tuning methods including LoRA and QLoRA for resource-efficient adaptation. Wren AI offers a semantic modeling layer on top of databases, allowing users to define business logic that guides SQL generation. Our system differs in its tight integration of PostgreSQL diagnostics into the self-healing loop, its dual API surface (native REST plus OpenAI-compatible), real-time repair-process streaming via Redis Pub/Sub, and single-command Docker Compose deployment.

\section{System Architecture}
\label{sec:architecture}

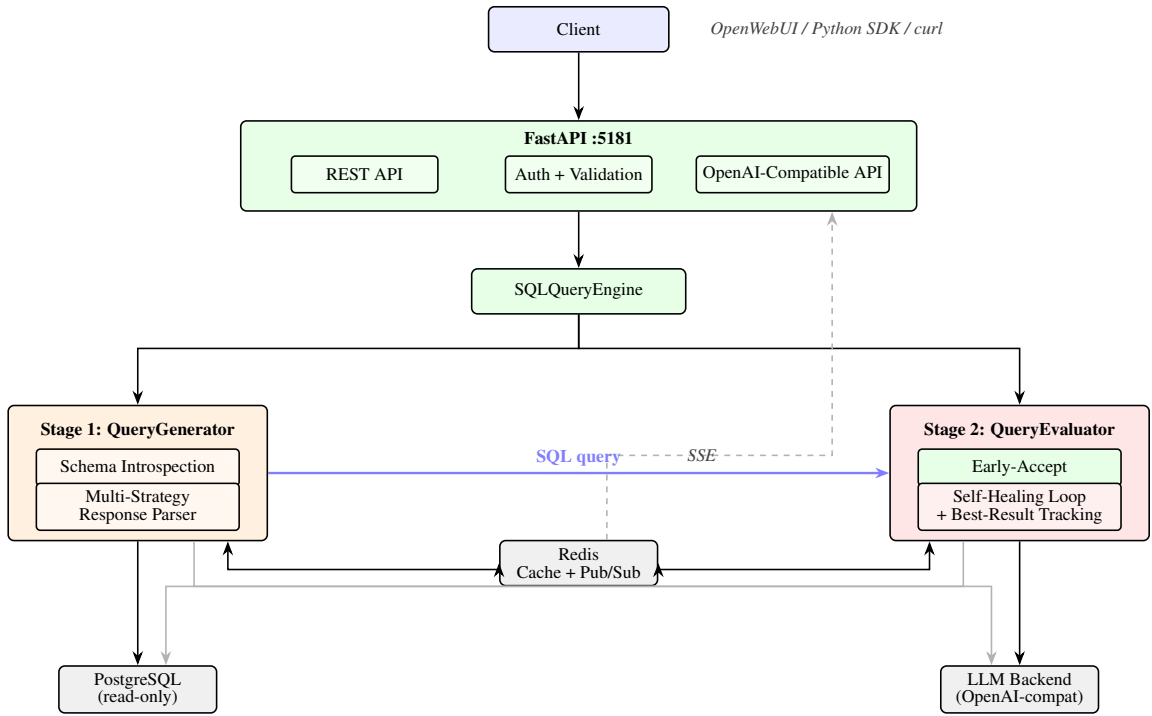
\begin{figure*}[ht]
\centering
\resizebox{0.92\textwidth}{!}{%
\begin{tikzpicture}[
  block/.style={draw, rounded corners=3pt, minimum height=0.8cm,
                align=center, font=\small, thick},
  subblock/.style={draw, rounded corners=2pt, minimum height=0.65cm,
                   align=center, font=\footnotesize, thick},
  ext/.style={draw, rounded corners=3pt, minimum width=2.8cm, minimum height=0.8cm,
              align=center, font=\small, thick, fill=gray!12},
  arr/.style={-{Stealth[length=2.5mm]}, thick},
  darr/.style={-{Stealth[length=2.5mm]}, thick, dashed, gray!60},
  biarr/.style={{Stealth[length=2.5mm]}-{Stealth[length=2.5mm]}, thick},
]

  \node[block, fill=blue!8, minimum width=3.2cm] (client) {Client};
  \node[font=\small\itshape, text=gray!50!black, right=0.6cm of client] {OpenWebUI / Python SDK / curl};

  \node[block, fill=green!10, below=1.2cm of client, minimum width=12.0cm, minimum height=1.6cm] (api) {};
  \node[font=\small\bfseries, anchor=north] at ([yshift=-0.08cm]api.north) {FastAPI :5181};
  \node[subblock, fill=green!6, minimum width=2.6cm] at ([xshift=-3.8cm, yshift=-0.15cm]api.center) (rest) {REST API};
  \node[subblock, fill=green!6, minimum width=2.6cm] at ([yshift=-0.15cm]api.center) (auth) {Auth + Validation};
  \node[subblock, fill=green!6, minimum width=3.2cm] at ([xshift=3.8cm, yshift=-0.15cm]api.center) (compat) {OpenAI-Compatible API};

  \node[block, fill=green!10, below=1.0cm of api, minimum width=3.8cm] (engine) {SQLQueryEngine};

  \node[block, fill=orange!12, below left=1.6cm and 3.6cm of engine,
        minimum width=4.6cm, minimum height=2.4cm] (gen) {};
  \node[font=\small\bfseries] at ([yshift=0.75cm]gen.center) {Stage 1: QueryGenerator};
  \node[subblock, fill=orange!6, minimum width=3.6cm] at ([yshift=0.1cm]gen.center) (introspect) {Schema Introspection};
  \node[subblock, fill=orange!6, minimum width=3.6cm] at ([yshift=-0.6cm]gen.center) (parser1) {Multi-Strategy\\[-1pt]Response Parser};

  \node[block, fill=red!10, below right=1.6cm and 3.6cm of engine,
        minimum width=4.6cm, minimum height=2.4cm] (eval) {};
  \node[font=\small\bfseries] at ([yshift=0.75cm]eval.center) {Stage 2: QueryEvaluator};
  \node[subblock, fill=green!10, minimum width=3.6cm] at ([yshift=0.1cm]eval.center) (earlyacc) {Early-Accept};
  \node[subblock, fill=red!6, minimum width=3.6cm] at ([yshift=-0.6cm]eval.center) (heal) {Self-Healing Loop\\[-1pt]+ Best-Result Tracking};

  \node[ext, below=2.2cm of gen] (pg) {PostgreSQL\\[-1pt](read-only)};
  \node[ext, below=4.0cm of engine] (redis) {Redis\\[-1pt]Cache + Pub/Sub};
  \node[ext, below=2.2cm of eval] (llm) {LLM Backend\\[-1pt](OpenAI-compat)};

  \draw[arr] (client) -- (api);
  \draw[arr] (api) -- (engine);
  \draw[arr] (engine.south) -- ++(0,-0.6) -| (gen.north);
  \draw[arr] (engine.south) -- ++(0,-0.6) -| (eval.north);

  \draw[arr] (gen) -- (pg);
  \draw[arr] (eval) -- (llm);

  \draw[arr, gray!60] ([xshift=1.0cm]gen.south) -- ++(0,-0.8) -| ([xshift=-0.5cm]llm.north);
  \draw[arr, gray!60] ([xshift=-1.0cm]eval.south) -- ++(0,-0.8) -| ([xshift=0.5cm]pg.north);

  \draw[biarr] ([xshift=1.6cm]gen.south) -- ++(0,-0.5) -| (redis.west);
  \draw[biarr] ([xshift=-1.6cm]eval.south) -- ++(0,-0.5) -| (redis.east);

  \draw[darr] ([xshift=0.5cm]redis.north) -- ++(0,1.5) -| ([xshift=4.5cm]api.south)
    node[pos=0.15, right=0.1cm, font=\footnotesize\itshape, text=gray!50!black] {SSE};

  \draw[arr, blue!50, very thick] (gen.east) -- node[above, font=\footnotesize\bfseries, text=blue!50] {SQL query} (eval.west);

\end{tikzpicture}%
}
\caption{Detailed architecture of SQL Query Engine. Stage~1 introspects the schema, generates SQL via the LLM, and extracts it through a five-strategy response parser (with \texttt{<think>} tag stripping for reasoning models). Stage~2 executes queries with early-accept for immediate success and a self-healing loop with best-result tracking for failures. Redis provides both session caching and Pub/Sub-based SSE streaming (dashed arrow). All PostgreSQL connections are read-only at the driver level.}
\label{fig:architecture}
\end{figure*}

Figure~\ref{fig:architecture} illustrates the high-level architecture. The system comprises three external dependencies (an OpenAI-compatible LLM endpoint, a PostgreSQL database, and a Redis instance) orchestrated by a FastAPI~\citep{fastapi2018} service containing six core modules.

\subsection{Module Responsibilities}

Table~\ref{tab:modules} summarizes the six core modules and their responsibilities.

\begin{table}[ht]
\centering
\caption{Core modules of SQL Query Engine.}
\label{tab:modules}
\small
\begin{tabular}{@{}lp{9.5cm}@{}}
\toprule
\textbf{Module} & \textbf{Responsibility} \\
\midrule
\texttt{engine.py}          & Orchestrates Stages 1 and 2; exposes \texttt{run()}, \texttt{generate()}, and \texttt{evaluate()} methods for full-pipeline, generation-only, and evaluation-only execution. \\
\texttt{queryGenerator.py}  & Stage 1: introspects the database schema via \texttt{dbHandler}, generates a detailed schema description through the LLM, caches it in Redis, and produces an initial SQL query extracted via a multi-strategy response parser. \\
\texttt{queryEvaluator.py}  & Stage 2: executes the query against PostgreSQL, captures errors, invokes the LLM evaluator, and iterates until success or retry exhaustion. Publishes progress to Redis Pub/Sub. \\
\texttt{dbHandler.py}       & Manages read-only PostgreSQL connections; provides schema introspection (\texttt{listTables}, \texttt{getTableSchema}, \texttt{getParsedSchemaDump}) and safe query execution. \\
\texttt{sessionManager.py}  & Redis-backed session store: serializes/deserializes LangChain~\citep{langchain2022} message objects, manages per-user context hashes, and maintains an auto-incrementing evaluation history counter. \\
\texttt{openaiCompat.py}    & Implements the OpenAI-compatible \texttt{/v1/chat/completions}, \texttt{/v1/completions}, and \texttt{/v1/models} endpoints with SSE streaming and Bearer token authentication. \\
\bottomrule
\end{tabular}
\end{table}

\subsection{Stage 1: Schema Introspection and SQL Generation}

On the first request for a given \texttt{chatID}, the \texttt{QueryGenerator} introspects the PostgreSQL database by querying \texttt{information\_schema.tables} and \texttt{information\_schema.columns}, then fetching a configurable number of sample rows per table. This raw schema, formatted as a markdown string, is passed to the LLM with a detailed system prompt that instructs it to produce a human-readable schema description covering table purposes, column semantics, relationships, JSONB structures, and data patterns. The resulting description is streamed to Redis Pub/Sub (enabling real-time visibility for clients) and cached in a Redis~\citep{redis2009} hash keyed by \texttt{\{chatID\}:SQLQueryEngine}.

Subsequent requests for the same \texttt{chatID} load the cached schema context directly, skipping introspection and LLM description generation entirely. The cached context includes the full LangChain message history, enabling multi-turn interactions in which the LLM has access to all prior questions and answers for the session.

The LLM's response is processed by a multi-strategy response parser (\texttt{\_parseResponse}), which attempts five extraction strategies in order: (1)~direct JSON parsing, (2)~embedded JSON extraction from surrounding text, (3)~code block extraction from markdown fences, (4)~\texttt{SELECT} regex matching, and (5)~raw text fallback. The parser targets an \texttt{AutomatedQuerySchema} with \texttt{description} and \texttt{query} fields, but field aliases (e.g., \texttt{sql} $\to$ \texttt{query}) are handled automatically via Pydantic model validators:

\begin{lstlisting}[style=pythonstyle, caption={Response schema for SQL generation with alias normalization.}]
class AutomatedQuerySchema(BaseModel):
    description: str = Field(
        ..., description="A description of the query.")
    query: str = Field(
        ..., description="The SQL query to execute.")
    # model_validator normalizes 'sql' -> 'query'
\end{lstlisting}

This design decouples the system from any single model's output format. It works reliably with models that produce JSON, markdown code blocks, or plain SQL text, without requiring structured output APIs or function calling. Additionally, \texttt{<think>} tags emitted by reasoning models (e.g., Qwen3, DeepSeek-R1) are stripped before parsing.

\subsection{Stage 2: Evaluation and Self-Healing}

The second stage receives the SQL query from Stage~1 (or from an external source, when called independently) and enters the self-healing loop described in Section~\ref{sec:selfhealing}. It resolves schema context through a three-tier priority system: (1)~\textbf{Payload}, context passed directly from Stage~1's output (zero-cost, no Redis round-trip); (2)~\textbf{Redis}, cached schema description from a prior call for the same \texttt{chatID}; (3)~\textbf{Scratch}, full database introspection and LLM-generated description (most expensive, used only as a last resort). This fallback chain ensures that the evaluator works correctly whether invoked immediately after generation or independently.

\section{The Self-Healing Algorithm}
\label{sec:selfhealing}

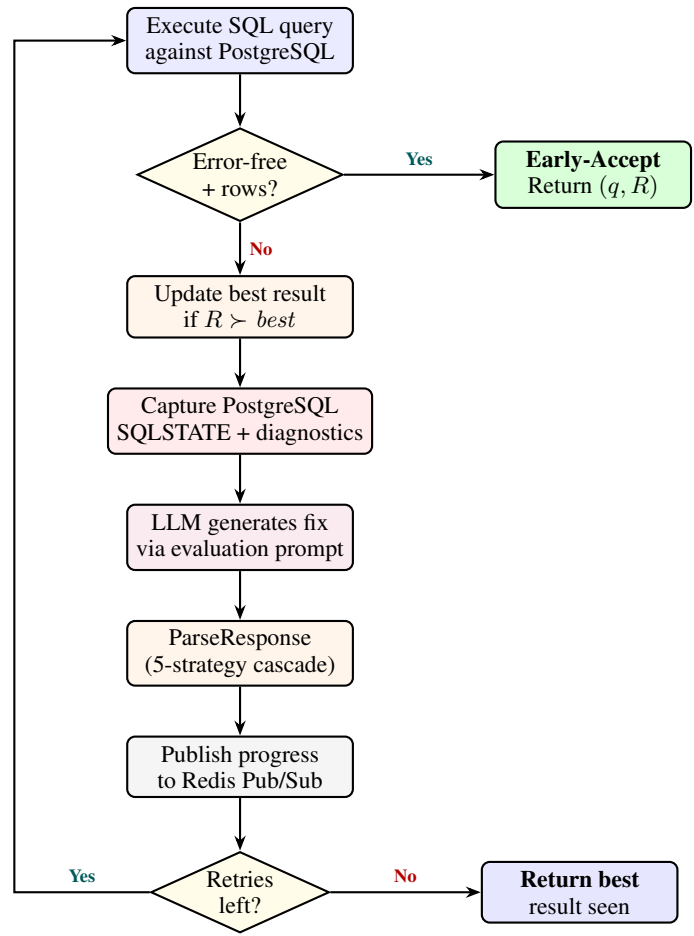
\begin{figure}[ht]
\centering
\begin{tikzpicture}[
  node distance=0.65cm,
  block/.style={draw, rounded corners=3pt, minimum width=3.0cm, minimum height=0.65cm,
                align=center, font=\footnotesize, thick},
  decision/.style={draw, diamond, aspect=2.2, minimum width=1.5cm, minimum height=0.5cm,
                   align=center, font=\footnotesize, thick, inner sep=1pt},
  result/.style={draw, rounded corners=3pt, minimum width=2.6cm, minimum height=0.65cm,
                 align=center, font=\footnotesize, thick},
  arr/.style={-{Stealth[length=2mm]}, thick},
  yes/.style={font=\scriptsize\bfseries, text=teal!70!black},
  no/.style={font=\scriptsize\bfseries, text=red!70!black},
]

  \node[block, fill=blue!8] (exec) {Execute SQL query\\against PostgreSQL};
  \node[decision, fill=yellow!10, below=0.7cm of exec] (success) {Error-free\\+ rows?};
  \node[result, fill=green!15, right=2.0cm of success] (accept) {\textbf{Early-Accept}\\Return $(q, R)$};
  \node[block, fill=orange!8, below=0.7cm of success] (track) {Update best result\\if $R \succ \mathit{best}$};
  \node[block, fill=red!8, below=0.65cm of track] (capture) {Capture PostgreSQL\\SQLSTATE + diagnostics};
  \node[block, fill=purple!8, below=0.65cm of capture] (llmfix) {LLM generates fix\\via evaluation prompt};
  \node[block, fill=orange!8, below=0.65cm of llmfix] (parse) {ParseResponse\\(5-strategy cascade)};
  \node[block, fill=gray!8, below=0.65cm of parse] (publish) {Publish progress\\to Redis Pub/Sub};
  \node[decision, fill=yellow!10, below=0.7cm of publish] (retry) {Retries\\left?};
  \node[result, fill=blue!10, right=2.0cm of retry] (best) {\textbf{Return best}\\result seen};

  \draw[arr] (exec) -- (success);
  \draw[arr] (success) -- node[yes, above] {Yes} (accept);
  \draw[arr] (success) -- node[no, right] {No} (track);
  \draw[arr] (track) -- (capture);
  \draw[arr] (capture) -- (llmfix);
  \draw[arr] (llmfix) -- (parse);
  \draw[arr] (parse) -- (publish);
  \draw[arr] (publish) -- (retry);
  \draw[arr] (retry) -- node[no, above] {No} (best);
  \draw[arr] (retry.west) -- node[yes, above] {Yes} ++(-1.8,0) |- (exec.west);

\end{tikzpicture}
\caption{Decision flow within the self-healing loop. The two key safety mechanisms are: \textbf{early-accept} (top right), which returns successful queries immediately without LLM re-evaluation; and \textbf{best-result tracking} (bottom right), which returns the best partial result when retries exhaust. Each iteration publishes progress to Redis Pub/Sub for real-time client visibility.}
\label{fig:selfhealing}
\end{figure}

Algorithm~\ref{alg:selfhealing} formalizes the self-healing loop, and Figure~\ref{fig:selfhealing} visualizes its decision flow.

\begin{figure}[ht]
\centering
\begin{tikzpicture}
\node[draw, rounded corners=4pt, fill=gray!4, inner sep=8pt, text width=0.92\textwidth] {
\small
\textbf{Algorithm 1:} \textsc{Self-Healing SQL Evaluation Loop}\\[6pt]
\textbf{Input:} candidate query $q_0$, user prompt $p_0$, schema context $S$, retry limit $N$, feedback row limit $k$\\
\textbf{Output:} best validated query $q^*$ with results $R^*$\\[6pt]
\begin{tabular}{@{}r@{\;}l@{}}
1: & $q \leftarrow q_0$;\; $p \leftarrow p_0$;\; $\mathit{obs} \leftarrow \text{``initial generation''}$;\; $\mathit{best} \leftarrow \textsc{None}$ \\[2pt]
2: & \textbf{for} $\;i = 1\;$ \textbf{to} $\;N\;$ \textbf{do} \\
3: & \quad $R, E \leftarrow \textsc{Execute}(q)$ \hfill \textcolor{gray}{\textit{// run query against PostgreSQL}} \\
4: & \quad $\textsc{Publish}(\texttt{chatID},\; q,\; \mathit{obs},\; E)$ \hfill \textcolor{gray}{\textit{// stream progress via Redis Pub/Sub}} \\[2pt]
5: & \quad \textbf{if} $\;E = \varnothing \;\wedge\; |R| > 0\;$ \textbf{then} \\
6: & \qquad \textbf{return} $\;(q, R)$ \hfill \textcolor{teal}{\textit{// \textbf{early-accept}: success without LLM re-check}} \\[2pt]
7: & \quad \textbf{if} $\;R \succ \mathit{best}\;$ \textbf{then} $\;\mathit{best} \leftarrow (q, R)$ \hfill \textcolor{teal}{\textit{// \textbf{best-result tracking}}} \\[2pt]
8: & \quad $q',\; p',\; \mathit{obs}' \leftarrow \textsc{LLM-Eval}(S,\; p,\; q,\; R[\,{:}k\,],\; E)$ \hfill \textcolor{gray}{\textit{// \texttt{isValid} field ignored}} \\
9: & \quad $\textsc{Publish}(\texttt{chatID},\; \mathit{obs}',\; q',\; p')$ \\
10: & \quad $q \leftarrow \textsc{ParseResponse}(q')$;\; $p \leftarrow p'$;\; $\mathit{obs} \leftarrow \mathit{obs}'$ \hfill \textcolor{gray}{\textit{// 5-strategy cascade}} \\[2pt]
11: & \textbf{end for} \\[2pt]
12: & \textbf{return} $\;\mathit{best}$ \hfill \textcolor{teal}{\textit{// return best seen, never \textsc{None}}} \\
\end{tabular}
};
\end{tikzpicture}
\caption{The self-healing evaluation loop. Key design choices are highlighted in \textcolor{teal}{teal}: early-accept (line~6) prevents regressions by never re-evaluating successful queries; best-result tracking (line~7) ensures graceful degradation on retry exhaustion (line~12). The \texttt{isValid} self-assessment is ignored (line~8); only execution determines success.}
\label{alg:selfhealing}
\end{figure}

Four design decisions warrant explanation.

\paragraph{Early-accept.} If a query executes without error and returns at least one row, it is accepted immediately; the LLM is never asked to re-evaluate a working result (lines~5--6). This prevents a critical failure mode in iterative refinement: the LLM ``fixing'' a correct query into an incorrect one. Our evaluation (Section~\ref{sec:evaluation}) confirms that models without this protection exhibit regressions.

\paragraph{Best-result tracking.} The loop tracks the best result seen across all iterations (line~7). If retries exhaust without producing a correct fix, the system returns the best partial result rather than failing (line~12). Together with early-accept, this guarantees the pipeline never downgrades a previously successful result.

\paragraph{Ignoring LLM self-assessment.} The \texttt{QueryEvaluationSchema} includes an \texttt{isValid} field, but it is always set to \texttt{False} by design. The system never trusts the LLM's judgment about query correctness; only execution outcomes (no error + non-empty result set) determine success. This eliminates a class of errors where the LLM incorrectly claims a broken query is valid.

\paragraph{Rich error context.} When a \texttt{psycopg.Error} is caught, the system extracts the SQLSTATE code (e.g., \texttt{42P01} for ``relation does not exist''), the primary diagnostic message, detail, and hint fields from PostgreSQL's diagnostic object~\citep{psycopg3}, as well as the full Python traceback. This level of detail allows the LLM to pinpoint the problem rather than guessing. The following is a representative error string passed to the LLM:

\begin{lstlisting}[basicstyle=\ttfamily\footnotesize, frame=single, breaklines=true]
Psycopg Error caught:
Error type: UndefinedColumn
Error message: column "order_date" does not exist
SQLSTATE code: 42703
PostgreSQL diag.message_hint: Perhaps you meant
  to reference the column "orders.order_timestamp".
\end{lstlisting}

\paragraph{Evaluation response parsing.} The LLM evaluator targets a \texttt{QueryEvaluationSchema} with four fields (\texttt{isValid}, \texttt{fixedQuery}, \texttt{observation}, \texttt{modifiedUserPrompt}), but the response is extracted via the same multi-strategy parser used in Stage~1 rather than a structured output API:

\begin{lstlisting}[style=pythonstyle, caption={Evaluation schema with alias normalization. The \texttt{isValid} field is always \texttt{False} by design.}]
class QueryEvaluationSchema(BaseModel):
    isValid: bool = Field(...)  # always False
    modifiedUserPrompt: str = Field(...)
    observation: str = Field(...)
    fixedQuery: str = Field(...)
    # model_validator normalizes aliases:
    # 'fixed_query' -> 'fixedQuery', etc.
\end{lstlisting}

Field aliases (e.g., \texttt{fixed\_query} $\to$ \texttt{fixedQuery}) are handled by a Pydantic model validator, ensuring compatibility across models that use different casing conventions.

\paragraph{Prompt modification.} The evaluator can optionally modify the user's original prompt (\texttt{modifiedUserPrompt}) to better align it with the discovered schema. For example, if the user asks about ``orders last month'' but the date column is named \texttt{transaction\_timestamp}, the evaluator might adjust the prompt to reference the correct column semantics. This modified prompt is carried forward into subsequent retry iterations.

\section{Real-Time Streaming Protocol}
\label{sec:streaming}

Users need to see what the repair process is doing. The system exposes real-time streaming through a two-layer architecture.

\paragraph{Layer 1: Redis Pub/Sub.} Every significant event in both stages is published to a Redis channel keyed by \texttt{chatID}. Messages follow a structured format:

\begin{lstlisting}[basicstyle=\ttfamily\footnotesize, frame=single, breaklines=true]
</{component}:{event}><|-/|-/>{content}
\end{lstlisting}

\noindent where the delimiter \texttt{<|-/|-/>} separates the event tag from the content payload. Representative events include \texttt{SQLQueryGenerator:schemaDescriptionChat} (streaming schema description chunks) and \texttt{SQLQueryEvaluator:QueryFixAttempt\#1} (evaluation progress for the first retry).

\paragraph{Layer 2: Server-Sent Events (SSE).} The OpenAI-compatible endpoint wraps the Redis Pub/Sub messages as SSE chunks conforming to the OpenAI streaming format~\citep{hickson2015sse}. Progress messages are enclosed in \texttt{<think>...</think>} tags, which reasoning-capable clients such as Open~WebUI render as chain-of-thought visualizations. The final result (SQL query plus a markdown-formatted results table) is emitted after the closing \texttt{</think>} tag.

The SSE subscription is established \emph{before} the engine is launched in a thread-pool executor, ensuring that no early messages are lost. A drain loop after engine completion collects any in-flight messages. The same content is simultaneously published to a secondary channel (\texttt{\{chatID\}:stream}) for external subscribers that do not consume SSE.

\section{Dual API Surface}
\label{sec:api}

The system exposes two complementary API surfaces.

\paragraph{Native REST API.} Three endpoints provide fine-grained control: \texttt{POST /inference/sqlQueryEngine/\{chatID\}} for the full pipeline, \texttt{POST /inference/sqlQueryGeneration/\{chatID\}} for Stage~1 only, and \texttt{POST /inference/sqlQueryEvaluation/\{chatID\}} for Stage~2 only. All LLM, PostgreSQL, and Redis connection parameters are declared as query parameters (with environment variable defaults), so they appear as individual labeled inputs in FastAPI's auto-generated Swagger UI.

\paragraph{OpenAI-compatible API.} The \texttt{/v1/chat/completions} and \texttt{/v1/completions} endpoints accept the standard OpenAI request schema and return responses in the standard OpenAI format, with both streaming and non-streaming modes. Any OpenAI-compatible client can connect without modification. The \texttt{/v1/models} endpoint lists the engine as a model. Authentication is handled via Bearer tokens, with support for multiple comma-separated keys.

When Open~WebUI injects a \texttt{chat\_id} field in the request body, that value is used directly as the Redis namespace key. When \texttt{chat\_id} is absent, a stable fallback is derived by MD5-hashing the first user message, providing a consistent session key across turns without requiring explicit session management from the client.

\section{Safety Mechanisms}
\label{sec:safety}

Running LLM-generated SQL against a live database in production demands multiple layers of safety.

\paragraph{Read-only enforcement.} The PostgreSQL connection is set to read-only mode at the \texttt{psycopg3} driver level immediately after connection establishment via \texttt{conn.set\_read\_only(True)}. This is a hard boundary enforced by the database driver: even if the LLM generates an \texttt{INSERT}, \texttt{UPDATE}, \texttt{DELETE}, or \texttt{DROP} statement, the database will reject it. This is strictly more reliable than prompt-based guardrails alone.

\paragraph{Result limiting.} A configurable hard limit (default: 50 rows) caps the number of result rows returned to the client, preventing memory exhaustion from runaway queries. The feedback loop uses a separate, smaller limit (\texttt{feedbackExamples}, default: 3 rows) to control how much data the LLM sees during evaluation, reducing token consumption.

\paragraph{Session isolation.} Each \texttt{chatID} maps to a dedicated Redis hash namespace (\texttt{\{chatID\}:SQLQueryEngine}). No data leaks between sessions. Evaluation histories are stored under auto-incrementing keys (\texttt{validatorChat:1}, \texttt{validatorChat:2}, etc.) to prevent overwrites across calls.

\paragraph{API authentication.} The OpenAI-compatible routes support Bearer token authentication via the \texttt{OPENAI\_API\_KEY} environment variable. Multiple keys can be comma-separated for multi-user deployments.

\paragraph{Transaction rollback.} After every query execution error, the system explicitly calls \texttt{conn.rollback()} to reset the transaction state, preventing error accumulation across retry iterations.

\section{Empirical Evaluation}
\label{sec:evaluation}

We evaluate the self-healing pipeline on two complementary benchmarks: a controlled synthetic suite that isolates the effect of the self-healing loop under known conditions, and the BIRD benchmark~\citep{li2024bird} that tests the system against real-world databases and queries at scale. Both evaluations use the same three-configuration ablation design across five LLM backends served via an OpenAI-compatible inference API.

\subsection{Common Experimental Protocol}

\paragraph{Configurations.} Three configurations isolate each pipeline stage. \textbf{Config~C} (\texttt{retryCount}${}=0$): generation only, where the LLM produces a single query with no evaluation or repair. \textbf{Config~B} (\texttt{retryCount}${}=1$): a single evaluation pass and one repair attempt. \textbf{Config~A} (\texttt{retryCount}${}=5$): the full pipeline with up to five repair iterations.

\paragraph{LLM backends.} Five models are evaluated: GPT-OSS-20B (20B parameters), Llama~3.3-70B (70B), GPT-OSS-120B (120B), Llama~4~Scout-17B (17B MoE with 16 experts), and Qwen3-32B (32B). All models are accessed through the same OpenAI-compatible endpoint, with temperature set to 0.1 and no system-level prompt modifications beyond the standard SQL generation and evaluation templates.\footnote{GPT-OSS-20B and GPT-OSS-120B refer to open-source models hosted under the identifiers \texttt{openai/gpt-oss-20b} and \texttt{openai/gpt-oss-120b} on the Groq inference platform. The remaining models use their canonical identifiers: \texttt{llama-3.3-70b-versatile}, \texttt{meta-llama/llama-4-scout-17b-16e-instruct}, and \texttt{qwen/qwen3-32b}.}

\paragraph{Evaluation metric.} A question is scored as correct if the engine's result rows, after normalization (case folding, whitespace trimming, numeric rounding to two decimal places, and column reordering), exactly match the gold query's result rows executed against the same database snapshot.

\paragraph{Regression counting.} A \emph{regression} is a question answered correctly under Config~C (generation only) but incorrectly under Config~A (full pipeline). Regressions indicate cases where the self-healing loop degraded a correct initial query, which is precisely the failure mode that early-accept is designed to prevent.

\subsection{Evaluation I: Synthetic Benchmark}
\label{sec:eval_synthetic}

\paragraph{Databases.} Three synthetic PostgreSQL databases are generated using Faker~\citep{faker2023}: an e-commerce database (customers, orders, products, reviews), a healthcare database (patients, providers, appointments, prescriptions), and a university database (students, courses, enrollments, departments). Each database contains realistic multi-table schemas with foreign keys, JSONB columns, and varied data types.

\paragraph{Questions.} 75 gold-standard questions (25 per database) span four difficulty tiers: \emph{easy} (single-table lookups), \emph{medium} (joins, aggregations), \emph{hard} (subqueries, CTEs, window functions), and \emph{extra hard} (multi-step reasoning, type casting, JSONB operators). Each question is paired with a verified gold SQL query.

\paragraph{Results.}

\begin{table}[ht]
\centering
\caption{Synthetic benchmark: execution accuracy (\%) across three configurations and five LLM backends (75 questions). \emph{Delta} is the change from Config~C to Config~A. \emph{Regr.}\ counts questions correct in Config~C but incorrect in Config~A.}
\label{tab:synthetic_results}
\small
\begin{tabular}{@{}lcccccc@{}}
\toprule
\textbf{Model} & \textbf{C} & \textbf{B} & \textbf{A} & \textbf{$\Delta$ (C$\to$A)} & \textbf{Regr.} \\
\midrule
GPT-OSS-20B                      & 45.3 & 45.3 & 53.3 & +8.0  & 0 \\
Llama~3.3-70B                    & 53.3 & 54.7 & 54.7 & +1.4  & 2 \\
GPT-OSS-120B                     & 50.7 & 49.3 & 48.0 & $-$2.7 & 4 \\
\textbf{Llama~4~Scout-17B}       & 48.0 & 50.7 & \textbf{57.3} & \textbf{+9.3} & \textbf{0} \\
Qwen3-32B                        & 48.0 & 42.7 & 46.7 & $-$1.3 & 6 \\
\bottomrule
\end{tabular}
\end{table}

\begin{table}[ht]
\centering
\caption{Synthetic benchmark: accuracy by difficulty tier under Config~A (\%). Easy questions (21) are near-saturated; the self-healing loop has its largest effect on hard queries.}
\label{tab:synthetic_difficulty}
\small
\begin{tabular}{@{}lcccc@{}}
\toprule
\textbf{Model} & \textbf{Easy} & \textbf{Medium} & \textbf{Hard} & \textbf{Extra Hard} \\
\midrule
GPT-OSS-20B          & 95.2 & 66.7 & 44.4 & 0.0 \\
Llama~3.3-70B        & 100.0 & 77.8 & 27.8 & 5.6 \\
GPT-OSS-120B         & 100.0 & 55.6 & 27.8 & 0.0 \\
Llama~4~Scout-17B    & 100.0 & 77.8 & 44.4 & 0.0 \\
Qwen3-32B            & 95.2 & 61.1 & 22.2 & 0.0 \\
\bottomrule
\end{tabular}
\end{table}

Tables~\ref{tab:synthetic_results} and~\ref{tab:synthetic_difficulty} present the synthetic results. Several findings are noteworthy.

\paragraph{Self-healing magnitude varies by model.} Llama~4~Scout shows the strongest response to iterative repair, gaining +9.3 percentage points from Config~C to Config~A with zero regressions. GPT-OSS-20B also benefits substantially (+8.0pp, zero regressions). In contrast, GPT-OSS-120B and Qwen3-32B exhibit slight degradation ($-$2.7pp and $-$1.3pp respectively), driven by regressions where the repair loop replaced correct-but-empty-result queries with incorrect alternatives.

\paragraph{Early-accept prevents regressions.} The two models with zero regressions (Llama~4~Scout, GPT-OSS-20B) are precisely those whose initially correct queries returned non-empty result sets, allowing early-accept to short-circuit the repair loop. Models with regressions (GPT-OSS-120B: 4, Qwen3-32B: 6) produced initially correct queries that happened to return empty results on the first execution, thereby entering the repair loop and suffering degradation.

\paragraph{Model size is not predictive.} The best-performing model (Llama~4~Scout at 17B MoE parameters) is substantially smaller than GPT-OSS-120B (120B dense parameters), suggesting that a model's capacity to leverage error diagnostics for self-correction is not merely a function of parameter count. The mixture-of-experts architecture may provide an advantage here, as different experts can specialize in different aspects of the repair task.

\paragraph{Difficulty ceiling.} All models achieve near-perfect accuracy on easy questions (95--100\%) but struggle with extra-hard queries (0--5.6\%). The self-healing loop has its largest absolute effect on hard-tier questions, where Llama~4~Scout improves from 5.6\% (Config~C) to 44.4\% (Config~A).

\subsection{Evaluation II: BIRD Benchmark}
\label{sec:eval_bird}

To validate the self-healing pipeline on real-world data at a larger scale, we evaluate on the BIRD benchmark~\citep{li2024bird}, which contains complex questions over large, messy databases drawn from professional domains.

\paragraph{Dataset.} We use the BIRD mini-dev split comprising 500 questions across 11 databases (california\_schools, card\_games, codebase\_community, debit\_card\_specializing, european\_football\_2, financial, formula\_1, student\_club, superhero, thrombosis\_prediction, and toxicology). Each question is paired with a gold SQL query and an optional ``evidence'' hint describing domain-specific knowledge; our evaluation runs \emph{without} evidence to measure the system's ability to operate from schema context alone.

\paragraph{SQLite-to-PostgreSQL migration.} BIRD's gold queries and databases use SQLite, whereas our system targets PostgreSQL. We implement an automated migration pipeline that applies 16 dialect-specific conversion rules covering data type mappings (\texttt{INTEGER PRIMARY KEY} $\to$ \texttt{SERIAL}), string functions (\texttt{SUBSTR} $\to$ \texttt{SUBSTRING}), date handling (\texttt{strftime} $\to$ \texttt{TO\_CHAR}/\texttt{EXTRACT}), type casting (\texttt{CAST(x AS REAL)} $\to$ \texttt{CAST(x AS DOUBLE PRECISION)}), and other syntactic differences (\texttt{GROUP\_CONCAT} $\to$ \texttt{STRING\_AGG}, \texttt{IFNULL} $\to$ \texttt{COALESCE}). Of the 500 questions, 63 (12.6\%) could not be converted automatically due to complex SQLite-specific constructs (nested \texttt{strftime} calls, non-standard \texttt{JULIANDAY} arithmetic, or ambiguous type coercions) and were excluded, leaving 437 evaluated questions.

\paragraph{Results.}

\begin{table}[ht]
\centering
\caption{BIRD benchmark: execution accuracy (\%) across three configurations and five LLM backends (437 questions after excluding 63 unconvertible queries). The self-healing loop yields up to +4.6pp improvement.}
\label{tab:bird_results}
\small
\begin{tabular}{@{}lcccccc@{}}
\toprule
\textbf{Model} & \textbf{C} & \textbf{B} & \textbf{A} & \textbf{$\Delta$ (C$\to$A)} & \textbf{Regr.} \\
\midrule
GPT-OSS-20B                      & 43.5 & 39.8 & 43.2 & $-$0.3 & 24 \\
Llama~3.3-70B                    & 43.7 & 38.0 & 46.5 & +2.8  & 23 \\
\textbf{GPT-OSS-120B}            & 44.4 & 42.6 & \textbf{49.0} & \textbf{+4.6} & 19 \\
Llama~4~Scout-17B                & 37.1 & 34.3 & 40.5 & +3.4  & 20 \\
Qwen3-32B                        & 40.7 & 41.6 & 39.4 & $-$1.3 & 17 \\
\bottomrule
\end{tabular}
\end{table}

\begin{table}[ht]
\centering
\caption{BIRD benchmark: accuracy by difficulty tier under Config~A (\%). BIRD uses three tiers: simple (284 questions), moderate (113), and challenging (40).}
\label{tab:bird_difficulty}
\small
\begin{tabular}{@{}lccc@{}}
\toprule
\textbf{Model} & \textbf{Simple} & \textbf{Moderate} & \textbf{Challenging} \\
\midrule
GPT-OSS-20B          & 54.9 & 42.3 & 25.0 \\
Llama~3.3-70B        & 55.6 & 47.9 & 26.2 \\
GPT-OSS-120B         & 59.9 & 48.8 & 30.0 \\
Llama~4~Scout-17B    & 49.3 & 41.4 & 22.5 \\
Qwen3-32B            & 47.2 & 40.0 & 23.8 \\
\bottomrule
\end{tabular}
\end{table}

Tables~\ref{tab:bird_results} and~\ref{tab:bird_difficulty} present the BIRD results.

\paragraph{Self-healing helps the strongest models most.} On BIRD, the model that benefits most from iterative repair is GPT-OSS-120B, gaining +4.6pp (from 44.4\% to 49.0\%). Llama~4~Scout also gains meaningfully (+3.4pp) despite starting from a lower baseline. This pattern differs from the synthetic benchmark, where Llama~4~Scout led; on the more complex BIRD queries, the larger model's broader knowledge base appears to be an advantage during repair.

\paragraph{Higher regression rates on real-world data.} All five models exhibit substantially more regressions on BIRD (17--24 per model) compared to the synthetic benchmark (0--6). This reflects the greater difficulty of BIRD's questions: many gold queries produce empty results on certain database states, causing the repair loop to trigger on correct queries more frequently. The early-accept mechanism still prevents regressions when queries return rows, but it cannot protect correct queries that genuinely produce empty result sets.

\paragraph{Config~B can underperform Config~C.} On BIRD, three of five models show Config~B accuracy \emph{below} Config~C (GPT-OSS-20B, Llama~3.3-70B, Llama~4~Scout). A single repair attempt appears insufficient to fix complex real-world queries and instead introduces noise. Config~A recovers because additional iterations give the model more chances to converge on a correct fix.

\paragraph{Comparison with published BIRD baselines.} The original BIRD paper~\citep{li2024bird} reports ChatGPT (GPT-3.5-Turbo) at 39.3\% and Claude~2 at 42.7\% execution accuracy on the development set without oracle knowledge evidence. Our best result of 49.0\% (GPT-OSS-120B, Config~A) exceeds both baselines, though direct comparison requires caution since we evaluate on the mini-dev split (437 of 500 questions after SQLite-to-PostgreSQL conversion) rather than the full development set. We note that current state-of-the-art systems such as CHASE-SQL~\citep{chasesql2024} achieve substantially higher accuracy on BIRD-dev (${\sim}$73\%) using task-specific fine-tuning, multi-path candidate generation, and preference-optimized selection; our contribution is the self-healing delta rather than absolute accuracy, as our system uses a single general-purpose LLM with no specialized training or schema linking.

\subsection{Latency and Resource Cost}

Table~\ref{tab:latency} reports the computational cost of self-healing across both benchmarks. On synthetic data, average query latency increases from 1.67--4.85\,s (Config~C, generation only) to 3.72--9.05\,s (Config~A, full pipeline), a $1.9$--$2.5\times$ overhead reflecting the additional LLM round-trips during repair iterations. Throughput correspondingly decreases from 23--64 questions/min to 12--30 questions/min. On BIRD, latencies are uniformly higher (5--11\,s for Config~C, 9--36\,s for Config~A) due to larger schemas and more complex queries requiring more repair iterations. Qwen3-32B exhibits the highest Config~A latency on BIRD (36.0\,s average, p99\,=\,180\,s) due to frequent timeouts during repair, suggesting that its reasoning-model architecture struggles with the iterative correction format. Peak memory remains stable across all configurations (62--107\,MB), confirming that the self-healing loop adds compute cost but not memory overhead.

\begin{table}[ht]
\centering
\caption{Latency and throughput cost of self-healing. Avg.\ latency in seconds; throughput in questions per minute (QPM); p90 latency for Config~A in seconds.}
\label{tab:latency}
\small
\begin{tabular}{@{}l cc cc cc@{}}
\toprule
& \multicolumn{2}{c}{\textbf{Avg.\ Latency (s)}} & \multicolumn{2}{c}{\textbf{Throughput (QPM)}} & \multicolumn{2}{c}{\textbf{Latency p90 (s)}} \\
\cmidrule(lr){2-3} \cmidrule(lr){4-5} \cmidrule(lr){6-7}
\textbf{Model} & C & A & C & A & C & A \\
\midrule
\multicolumn{7}{@{}l}{\textit{Synthetic (75 questions)}} \\[2pt]
GPT-OSS-20B    & 1.72 & 3.72  & 64.3 & 30.2  & --- & 9.8 \\
Llama 3.3-70B  & 2.44 & 4.12  & 44.7 & 27.2  & --- & 10.0 \\
GPT-OSS-120B   & 2.83 & 7.00  & 37.9 & 16.1  & --- & 20.7 \\
\textbf{Scout-17B} & 1.67 & 4.19 & 63.8 & 27.0 & --- & 9.2 \\
Qwen3-32B      & 4.85 & 9.05  & 23.2 & 12.6  & --- & 23.5 \\
\midrule
\multicolumn{7}{@{}l}{\textit{BIRD (437 questions)}} \\[2pt]
GPT-OSS-20B    & 11.31 & 15.45 & 8.6  & 7.7   & --- & 19.1 \\
Llama 3.3-70B  & 4.95  & 9.06  & 23.0 & 12.8  & --- & 19.1 \\
\textbf{GPT-OSS-120B} & 5.11 & 8.72 & 22.3 & 13.3 & --- & 16.6 \\
Scout-17B      & 5.14  & 9.23  & 22.5 & 12.7  & --- & 18.7 \\
Qwen3-32B      & 7.20  & 36.01 & 16.0 & 3.3   & --- & 180.1 \\
\bottomrule
\end{tabular}
\end{table}

\begin{figure}[ht]
\centering

\small
\tikz\fill[blue!35] (0,0) rectangle (0.3,0.2); Config~C (generation only)
\qquad
\tikz\fill[teal!55] (0,0) rectangle (0.3,0.2); Config~A (full self-healing)

\vspace{4pt}
\resizebox{\textwidth}{!}{%
\begin{tikzpicture}[
  x=1.8cm, y=0.075cm,
]
  \node[font=\small\bfseries] at (2.6,65) {Synthetic (75 questions)};
  \node[font=\small\bfseries] at (8.6,65) {BIRD (437 questions)};

  \draw[thick, -{Stealth[length=2.5mm]}] (0,0) -- (0,62);
  \node[rotate=90, font=\small, anchor=south] at (-0.35,30) {Accuracy (\%)};
  \draw[thick] (0,0) -- (11.4,0);
  \foreach \y in {0,10,20,30,40,50,60} {
    \draw[gray!20] (0,\y) -- (11.4,\y);
    \draw (0,\y) -- (-0.05,\y) node[left, font=\small] {\y};
  }

  \fill[blue!35] (0.40,0) rectangle (0.72,45.3);
  \fill[teal!55] (0.80,0) rectangle (1.12,53.3);
  \node[font=\small, above] at (0.56,45.3) {45};
  \node[font=\small, above] at (0.96,53.3) {53};
  \node[font=\scriptsize\sffamily, below, text width=1.4cm, align=center] at (0.76,-2.5) {GPT-OSS\\[-1pt]20B};

  \fill[blue!35] (1.40,0) rectangle (1.72,53.3);
  \fill[teal!55] (1.80,0) rectangle (2.12,54.7);
  \node[font=\small, above] at (1.56,53.3) {53};
  \node[font=\small, above] at (1.96,54.7) {55};
  \node[font=\scriptsize\sffamily, below, text width=1.4cm, align=center] at (1.76,-2.5) {Llama\\[-1pt]3.3-70B};

  \fill[blue!35] (2.40,0) rectangle (2.72,50.7);
  \fill[teal!55] (2.80,0) rectangle (3.12,48.0);
  \node[font=\small, above] at (2.56,50.7) {51};
  \node[font=\small, above] at (2.96,48.0) {48};
  \node[font=\scriptsize\sffamily, below, text width=1.4cm, align=center] at (2.76,-2.5) {GPT-OSS\\[-1pt]120B};

  \fill[blue!35] (3.40,0) rectangle (3.72,48.0);
  \fill[teal!55] (3.80,0) rectangle (4.12,57.3);
  \node[font=\small, above] at (3.56,48.0) {48};
  \node[font=\small\bfseries, above, text=teal!70!black] at (3.96,57.3) {57};
  \node[font=\scriptsize\sffamily, below, text width=1.4cm, align=center] at (3.76,-2.5) {Llama~4\\[-1pt]Scout};

  \fill[blue!35] (4.40,0) rectangle (4.72,48.0);
  \fill[teal!55] (4.80,0) rectangle (5.12,46.7);
  \node[font=\small, above] at (4.56,48.0) {48};
  \node[font=\small, above] at (4.96,46.7) {47};
  \node[font=\scriptsize\sffamily, below, text width=1.4cm, align=center] at (4.76,-2.5) {Qwen3\\[-1pt]32B};

  \draw[gray!40, dashed, thick] (5.7,0) -- (5.7,62);

  \fill[blue!35] (6.40,0) rectangle (6.72,43.5);
  \fill[teal!55] (6.80,0) rectangle (7.12,43.2);
  \node[font=\small, above] at (6.56,43.5) {44};
  \node[font=\small, above] at (6.96,43.2) {43};
  \node[font=\scriptsize\sffamily, below, text width=1.4cm, align=center] at (6.76,-2.5) {GPT-OSS\\[-1pt]20B};

  \fill[blue!35] (7.40,0) rectangle (7.72,43.7);
  \fill[teal!55] (7.80,0) rectangle (8.12,46.5);
  \node[font=\small, above] at (7.56,43.7) {44};
  \node[font=\small, above] at (7.96,46.5) {47};
  \node[font=\scriptsize\sffamily, below, text width=1.4cm, align=center] at (7.76,-2.5) {Llama\\[-1pt]3.3-70B};

  \fill[blue!35] (8.40,0) rectangle (8.72,44.4);
  \fill[teal!55] (8.80,0) rectangle (9.12,49.0);
  \node[font=\small, above] at (8.56,44.4) {44};
  \node[font=\small\bfseries, above, text=teal!70!black] at (8.96,49.0) {49};
  \node[font=\scriptsize\sffamily, below, text width=1.4cm, align=center] at (8.76,-2.5) {GPT-OSS\\[-1pt]120B};

  \fill[blue!35] (9.40,0) rectangle (9.72,37.1);
  \fill[teal!55] (9.80,0) rectangle (10.12,40.5);
  \node[font=\small, above] at (9.56,37.1) {37};
  \node[font=\small, above] at (9.96,40.5) {41};
  \node[font=\scriptsize\sffamily, below, text width=1.4cm, align=center] at (9.76,-2.5) {Llama~4\\[-1pt]Scout};

  \fill[blue!35] (10.40,0) rectangle (10.72,40.7);
  \fill[teal!55] (10.80,0) rectangle (11.12,39.4);
  \node[font=\small, above] at (10.56,40.7) {41};
  \node[font=\small, above] at (10.96,39.4) {39};
  \node[font=\scriptsize\sffamily, below, text width=1.4cm, align=center] at (10.76,-2.5) {Qwen3\\[-1pt]32B};

\end{tikzpicture}%
}
\caption{Execution accuracy before (Config~C) and after (Config~A) the self-healing loop on both benchmarks. Scout leads on synthetic data (+9.3pp), GPT-OSS-120B leads on BIRD (+4.6pp).}
\label{fig:latency_comparison}
\end{figure}
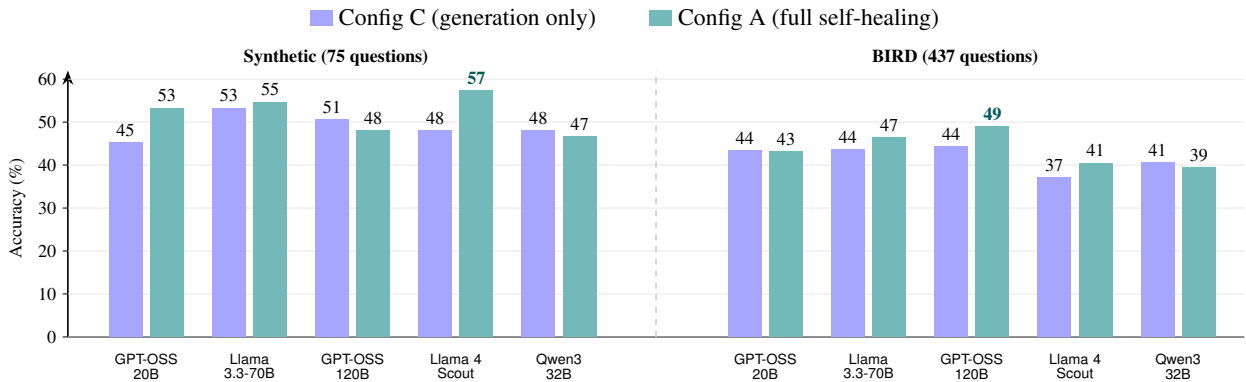

\subsection{Cross-Benchmark Analysis}

Comparing results across the two benchmarks reveals several patterns.

\paragraph{Different models excel on different tasks.} Llama~4~Scout is the top performer on synthetic data (57.3\%) but ranks fourth on BIRD (40.5\%). Conversely, GPT-OSS-120B is the weakest self-healer on synthetic data ($-$2.7pp delta) but the strongest on BIRD (+4.6pp). This suggests that the synthetic benchmark, with its cleaner schemas and more formulaic questions, rewards a model's ability to follow repair instructions precisely, while BIRD's messy real-world data rewards broader domain knowledge.

\paragraph{Self-healing effect size scales with task difficulty.} On synthetic easy-tier questions (where baseline accuracy is 95--100\%), self-healing has negligible effect because there is little room to improve. On BIRD's challenging tier (where baseline accuracy is 22--28\%), the absolute improvement is modest but the \emph{relative} improvement can reach 30\% or more. The loop is most valuable in the ``reachable failure'' zone: queries that are close to correct but need specific adjustments.

\paragraph{Common error patterns.} Across both benchmarks and all models, the most frequent error categories are: (1)~schema mismatches, where the LLM hallucinates column or table names not present in the database; (2)~type casting failures, particularly PostgreSQL's requirement for explicit \texttt{NUMERIC} casts in \texttt{ROUND} calls; (3)~CTE syntax errors, including incorrect \texttt{WITH} clause placement and recursive CTE misuse; and (4)~overly restrictive filters, especially date ranges and string matching conditions that produce empty results.

\subsection{Limitations}

The synthetic benchmark uses 75 questions across three purpose-built databases; the resulting confidence intervals are approximately $\pm$6pp at the 95\% level. The BIRD evaluation excludes 63 of 500 questions (12.6\%) due to SQLite-to-PostgreSQL conversion failures, which may introduce a bias toward questions with simpler SQL constructs. We do not evaluate on Spider due to its SQLite-native design; a PostgreSQL adaptation of Spider would strengthen cross-benchmark comparison. Finally, all models are accessed through a single inference provider, so results may differ under alternative serving configurations or quantisation levels.

\section{Discussion}
\label{sec:discussion}

\paragraph{Schema caching trade-offs.} Caching the LLM-generated schema description in Redis cuts latency substantially for subsequent queries in the same session (eliminating both database introspection and an LLM generation call). However, if the underlying database schema changes during a session, the cached description becomes stale. The system exposes a \texttt{schemaDescriptionKey} parameter that can be changed to force regeneration, and clearing the Redis key achieves the same effect. An automatic invalidation mechanism (e.g., listening to PostgreSQL DDL events) would be a useful extension.

\paragraph{Empty-result heuristic.} Treating empty result sets as failures is a design choice that biases the system toward producing queries that return data. This is appropriate for most analytical questions (``How many orders last month?'') but may produce unnecessary retries for existence queries (``Are there any orders from Antarctica?''). The LLM's evaluation prompt partially mitigates this by distinguishing between ``questions that expect data'' and ``questions where empty is acceptable,'' but the heuristic remains imperfect.

\paragraph{LLM backend flexibility.} Because the system communicates with any OpenAI-compatible endpoint, users can deploy it with local models via Ollama (e.g., Qwen~2.5-Coder at 7B parameters) for cost-free, air-gapped operation, or with hosted APIs (OpenAI, Anthropic via LiteLLM) for maximum quality. The trade-off between query quality and inference cost is left to the operator.

\paragraph{Prompt engineering as a configurable layer.} The system's SQL guidelines corpus (\texttt{sqlGuidelines.py}) injects several hundred lines of PostgreSQL best practices (covering JSONB operators, case-insensitive matching, window functions, CTEs, type casting, and common pitfalls) into every LLM prompt. Separate guideline sets are used for generation (\texttt{postgreManualData}) and evaluation (\texttt{postgreManualDataEval}), reflecting the different requirements of each stage. This corpus acts as a form of domain-specific retrieval augmentation, providing the LLM with specialized knowledge that improves query quality without requiring fine-tuning.

\paragraph{Comparison with multi-agent approaches.} Recent work has explored multi-agent architectures for text-to-SQL, where separate agents handle schema linking, query generation, and validation. Our system achieves a similar separation of concerns through its two-stage pipeline but uses a single LLM instance with different prompts for each stage, reducing infrastructure complexity.

\paragraph{Early-accept as a safety mechanism.} Our evaluation demonstrates that the early-accept design is not merely an optimization but a safety requirement. On the synthetic benchmark, models without regressions (Llama~4~Scout, GPT-OSS-20B) are precisely those whose correct queries returned non-empty results, allowing early-accept to short-circuit the loop. On BIRD, where many gold queries produce empty results, regression rates rise to 17--24 per model, confirming that the residual vulnerability lies in the empty-result heuristic rather than in the repair logic itself.

\paragraph{Limitations.} The system currently supports PostgreSQL only. Extending to other SQL dialects (MySQL, SQLite, Snowflake) would require dialect-specific guideline corpora and minor changes to the database handler. The self-healing loop adds latency proportional to the number of retries; a timeout mechanism (in addition to the retry count limit) would benefit latency-sensitive applications. The BIRD evaluation excludes 12.6\% of questions due to SQLite-to-PostgreSQL conversion failures, which may bias the results toward simpler SQL constructs. Finally, the system does not support multi-database queries or cross-database joins, though the modular architecture could accommodate such an extension.

\section{Deployment}
\label{sec:deployment}

The system is distributed as a Docker Compose~\citep{docker2013} stack with three services: the FastAPI application (port 5181), Redis (port 6380), and an optional Open~WebUI instance (port 5182) pre-configured to connect to the engine. A single \texttt{docker compose up -{}-build} command brings up the entire stack. All connection parameters are configured through environment variables in \texttt{docker-compose.yml}.

For programmatic use, the \texttt{SQLQueryEngine} class can be imported directly as a Python module, bypassing the HTTP layer entirely. Listing~3 demonstrates a minimal integration: the caller supplies LLM, database, and Redis connection parameters at construction time, then issues natural-language queries via the \texttt{run()} method.

\begin{lstlisting}[style=pythonstyle, caption={Using SQL Query Engine as a Python module.}]
from sqlQueryEngine import SQLQueryEngine

engine = SQLQueryEngine(
    llmParams={"model": "qwen2.5-coder:7b",
               "temperature": 0.1,
               "base_url": "http://localhost:11434/v1",
               "api_key": "ollama"},
    dbParams={"host": "localhost", "port": 5432,
              "dbname": "mydb", "user": "postgres",
              "password": "secret"},
    redisParams={"host": "localhost", "port": 6379,
                 "password": "", "db": 0,
                 "decode_responses": True}
)

result = engine.run(
    chatID="user123",
    basePrompt="How many orders were placed last month?"
)
\end{lstlisting}

\section{Conclusion}
\label{sec:conclusion}

This report has described SQL Query Engine, an open-source system that translates natural language into validated PostgreSQL queries through a two-stage, self-healing LLM pipeline. The main contributions are an iterative repair loop with early-accept and best-result tracking (up to +9.3pp accuracy gains on synthetic benchmarks, +4.6pp on BIRD, zero regressions on the best synthetic model), a multi-strategy response parser that handles any LLM output format without structured output APIs, a session-aware Redis caching layer that avoids redundant schema introspection across turns, a real-time streaming protocol that surfaces the repair process to clients, and a dual API surface compatible with existing OpenAI tooling. Our experiments across five LLM backends on both synthetic and real-world (BIRD) benchmarks confirm that execution-grounded self-healing is a practical technique for improving text-to-SQL accuracy, though the magnitude of improvement hinges on the model's capacity for iterative repair and on how often correct queries happen to return empty result sets. Read-only database access is enforced at the driver level, giving the system a hard safety boundary suitable for production deployment. Source code is available at \url{https://github.com/codeadeel/sqlqueryengine}.

Future work includes extending support to additional SQL dialects, integrating automatic schema change detection, adapting the Spider benchmark for PostgreSQL evaluation, and exploring multi-agent architectures~\citep{macsql2023} for more complex analytical queries that may benefit from specialized sub-agents for schema linking and query decomposition.

\section*{Acknowledgments}

The author thanks Dr.\ Spyridon Mastorakis for valuable discussions on system architecture and evaluation methodology during the early stages of this work.

\bibliographystyle{unsrtnat}
\bibliography{references}

\end{document}